\documentclass[11pt]{amsart}
\usepackage{geometry}                % See geometry.pdf to learn the layout options. There are lots.
\usepackage{graphicx}
\usepackage{amssymb}
\usepackage{epstopdf}
\usepackage{hyperref}
\usepackage{color}
\usepackage[shortlabels]{enumitem}
  \setlist[enumerate, 1]{1.}
\title{Typicality in the foundations of statistical physics and Born's rule} 
\author{Detlef D\"urr}
\address{Mathematisches Institut, Ludwig-Maximilians-Universitat M\"{u}nchen, Theresienstr. 39, D-80333 M\"{u}nchen, Germany}

\author{Ward Struyve}
\address{Instituut voor Theoretische Fysica \& Centrum voor Logica en Analytische Wijsbegeerte, KU Leuven, Belgium}
\begin{document}
\begin{abstract} 
Typicality has always been in the minds of the founding fathers of probability theory when probabilistic reasoning is applied to the real world. However, the role of typicality is not always appreciated. An example is the paper ``Foundations of statistical mechanics and the status of Born's rule in de Broglie-Bohm pilot-wave theory" by Antony Valentini \cite{V}, where he presents typicality and relaxation to equilibrium as {\em distinct} approaches to the proof of Born's rule, while typicality is in fact an overriding necessity. Moreover the ``typicality approach" to Born's rule of ``the Bohmian mechanics school'' is claimed  to be inherently circular. We wish to explain once more in very simple terms why the accusation is off target and why ``relaxation to equilibrium" is neither necessary nor sufficient to justify Born's rule.

 \end{abstract}

\maketitle
\noindent \sl Nino Zangh{\`i} and D.D.\ remember vividly the discussions with GianCarlo Ghirardi on Boltzmann's insights into statistical physics and its relation to the random theory he himself had proposed (with his coworkers) and had worked on for many decades until his untimely death. Not only was GianCarlo an admirer of Boltzmann, he also had a full grasp of Boltzmann's ideas and on the role of typicality. The GRW theory is intrinsically random and the $|\psi|^2$-distribution arises from the collapse mechanism built into the theory and he understood that the appeal to typicality, for empirical assertions, cannot be avoided.  We miss GianCarlo Ghirardi, our invaluable friend, coworker and colleague and we dedicate this work in memoriam to him.\rm

\section{Why ``most"  cannot be avoided} 
Typicality has always been in the minds of the founding fathers of probability theory when probabilistic reasoning is applied to the real world.  Nevertheless, still it's role is often not understood. An example is  \cite{V}, where Valentini  presents typicality and relaxation to equilibrium as {\em distinct} approaches to the proof of Born's rule, while typicality is in fact an overriding necessity. Valentini writes in the abstract of his article:
\begin{quote}
\emph{We compare and contrast two distinct approaches to understanding the Born rule in de Broglie-Bohm pilot-wave theory, one based on dynamical relaxation over time (advocated by this author and collaborators) and the other based on typicality of initial conditions (advocated by the `Bohmian mechanics' school). It is argued that the latter approach is inherently circular and physically misguided.}
\end{quote}
The accusation of circularity concerns the proof of Born's rule in de Broglie-Bohm pilot-wave theory, or ``Bohmian mechanics'' for short, given in \cite{DGZ}. It is an important proof, as it explains the observed regularity concerning the outcomes of measurements on ensembles of identically prepared systems. As such, Valentini's accusation is at the same time an onslaught to the ideas underlying statistical physics.
We wish  to explain  once more in very simple terms why the accusation is off target and why ``relaxation to equilibrium" is neither necessary nor sufficient to justify Born's rule. 

In the history of mathematics pointing out circularities in important proofs have been  sometimes pathbreaking. An example is provided by what we would call now the ``PhD thesis" of Georg Simon Kl{\"u}gel  (1739 - 1812), who showed that all existing proofs (about 27 of them) of the 11th Postulate of Euclid on the uniqueness of parallels were circular in that they used in the proofs equivalents of the postulate as (hidden) assumptions. That thesis has led to the discovery of non-Euclidean geometry! 

The accusation of circularity in the derivation of Born's rule is however less breathtaking; it is simply off target. The criticism misses the point of statistical physics entirely. That may be partly due to the loose manner of speaking about probability and distributions which is common in statistical physics and which clouds the meaning of these objects. Instead, the notion of typicality is necessary to understand what the statistical predictions of a physical theory really mean.  In fact, typicality (though the word  may not have been directly used) has always been in the minds of probabilists and physicists (from Jacob Bernoulli $\sim$ 1700 over Ludwig Boltzmann $\sim 1850$ to Kolmogorov's axiomatics of probability $\sim$ 1930 ) when probabilistic reasoning is applied to the real world. We try to explain once more in simplest terms why this need be. 

Let's start with a simple example: A (fair) coin is thrown, say, a thousand times (the number thousand is chosen only because it is kind of large). We obtain a head-tail sequence of length $1000$ and ask prior to inspection of the sequence: Roughly how many heads are there in the sequence? Some would perhaps prefer to be agnostic about the answer but most would say -- perhaps after some time of reflection -- that the number will be roughly 500. Actually they will find out by counting the heads that they were right. Why roughly 500? Well, the relative frequency of heads (or tails) is then $1/2$, the probability\footnote{Luckily, for what we have to say here, we can ignore the issues related to the question what the notion of ``probability $1/2$" \emph{really} means. It does not matter.} which we naturally assign for the sides of a (fair) coin. What matters here is that there is obviously some relation between the factual occurrence of the relative frequency of heads and the number $1/2$. That needs to be explained. Why? Because other sequences are possible as well, for example sequences which show less than $300$ heads. The question which needs to be answered is: Why don't they show up in practice? The (mathematical) way that the regularity of roughly $500$ heads is explained is by the \emph{law of large numbers (LLN)}, which establishes the closeness of the empirical distribution of heads, i.e., the distribution which counts the relative frequency of heads in the sequence of length $1000$ (which is the large number in the LLN) and the number $1/2$.
 
How does the LLN explain that? By counting sequences! Here are some telling numbers: There are about $10^{300}$ sequences with about $500$ heads. There are about $10^{260}$ sequences with about $300$ heads. So the proportion of sequences with $500$ heads versus $300$ heads is about $10^{40}$. For sharpening our intuition about the power of such numbers note that  the age of the universe in seconds is about $10^{17}$. Thus most\footnote{``Most'' should be understood as ``overwhelmingly many''.} sequences show a law-like regularity, namely that the relative number of heads is roughly $1/2$.  Wouldn't that suggest that it is most likely that the observed sequence has roughly equal numbers of heads and tails?  Well, most likely is just another way of saying ``with high probability". But then, what does probability mean here? It is better and simpler  to say that the \emph{typical} sequences will have roughly $500$ heads. The LLN says nothing more than that. It is a typicality statement. We remark for later that in introductory courses to probability theory the counting is introduced as Laplace probability which is then a normalized quantity by dividing the numbers of sequences of interest by the total number of sequences and which we better refer to by the role it plays in our example as Laplace-typicality. The point of this example is that mere counting of head-tail sequences (or $0-1$ sequences if one wants) gives two insights (where the second one we take it as being intuitively clear without going into details):
 \begin{enumerate}
 \item Most sequences show the law-like regularity that the empirical distribution of heads (the relative frequency of heads) is near $1/2$.
 \item The succession of heads and tails (or $1's$ and $0's$) in a typical sequence looks random, unpredictable, while randomness never entered. It's just the way typical sequences look like.
 \end{enumerate}
Agnostics may still complain that that explains nothing: What needs to be explained is why we only see typical sequences! That's actually the deep question underlying the meaning of probability theory from its very beginning and Antoine-Augustin Cournot (1801-1877) coined once what became known as Cournot's principle, which in our own rough words just says that we should only be concerned with typical events. The  point we wish to make with the simple example  is that appeal to typicality cannot be avoided.  Sequences with drastically unequal  number of heads and tails are physically possible. The reason they do not appear in practice is because there are much too few of them, they are atypical. There is no way around that. That is what the founders of probability theory understood and had to swallow. Note that typicality, through Cournot's principle, only tells us what to expect or not. It does not allow to associate a probability to, say, the sequence with 300 heads: In terms of the Laplace-typicality only values near zero or one matter -- atypical or typical. The notion of typicality is distinct from the notion of probability. 

Let's go a step further and consider the coin tossing as a physical process, because that is what it is after all: There is a hand which tosses the coin, thereby providing the coin with an initial momentum and position which determine its flight through the air. The trajectory of the coin is determined, given the initial conditions, by the laws of physics (here Newtonian) and hence it defines a function which maps initial conditions to head or tail ($0$ or $1$). But the hand is just a physical system itself -- a coin tossing machine so to say. The machine picks up the coin, throws it, and after the landing the machine notes down head or tail, picks the coin up again, throws it and so on and so forth. Thus the resulting sequence of heads and tails depends only on the initial conditions, i.e., on the phase space point which determines the whole process of the coin tossing machine. The physical description and analysis may not be that easy, but at least the principle is clear\footnote{For more on this, see the chapter ``Chance in Physics" in \cite{DT}.}: It shows that the head-tail-sequences are the images of a function $F=(F_1,\ldots, F_N)$ of the high dimensional phase space\footnote{Dimension of the size of Avogadro's number perhaps.} variables $q$ -- the initial conditions. Here the component $F_k$ maps to the outcome $\delta\in\{0,1\}$ of the $k$th tossing of the machine.  Such a function $F$ (a coarse-graining function by the way) is usually called a random variable. The point is that in this description where phase space variables play the decisive role, counting is not anymore possible, as classical phase space is a continuum. What then replaces the counting? That is a measure --  a \emph{typicality measure}. In classical physics, which would be appropriate for studying coin tossing as a physical process, the measure commonly used is the ``Liouville-measure" -- the volume measure in phase space. It recommends itself by the property of being stationary,\footnote{In classical mechanics, there are many more measures which share this property, but that does not matter for our concerns here.} an observation which was promoted in the works of Ludwig Boltzmann. It is a measure which is suggested by the physical law itself and not by an arbitrary human choice. The fact that, with this measure, typicality is a timeless notion is of great help for proving the LLN.

The role of the Liouville-measure is to define the notion of ``most" for the phase space of classical mechanics. In mathematical terms, the above mentioned Laplace-typicality emerges then ideally as the \emph{image measure} of the more fundamental Liouville-measure defined by the function $F$. To express the LLN in this more fundamental setting, it is useful to introduce the empirical distribution $\rho^N_{\rm emp}(q, \delta)$, the function which counts the relative number of heads and tails and which is a function of the phase space variables $q$ and the image variables $\delta\in\{0,1\} $.  $$\rho^N_{\rm emp}(q, \delta)= \frac{1}{N}\sum_{k=1}^N\mathbf{1}_{\{\delta\}}(F_k(q))\,.$$ Here $\mathbf{1}_{\{\delta\}}(F_k(q))=1$ if $F_k(q)=\delta$ and $0$ otherwise. The LLN (if it would be proven for a physically realistic coin tossing machine)  would then say something like:\footnote{A technical remark on the side: To model the coin tossing experiment in which the coin is thrown a great number of times in a physically realistic way is not so easy and to prove the LLN may turn out hard: The stochastic independence of the different tosses of the coin is easily said, but to prove that in a physically realistic model is far from being easy (see \cite{DT}, chapter ``Chance in Physics" for an elaboration on that).} 
 
 \medskip
\sl
\noindent For most phase points $q$ (w.r.t.\ the Liouville-measure) and for large enough $N$ the empirical distribution $\rho^N_{\rm emp}\approx 1/2$,  or,  Liouville-measure typically $\rho^N_{\rm emp}\approx 1/2$. 
\rm
\medskip

The reference to typicality cannot be avoided, as there are phase points which are mapped to sequences with, say, $300$ heads, i.e., ``most" cannot be replaced by ``all".

One further point should be noted which is often used to actually justify the use of statistical methods in physics: It is almost impossible to know in a realistic physical system exactly which initial conditions lead to which outcomes (as for example in the case of the coin tossing machine). The power of typicality is that exact details are not needed. It suffices that for most initial conditions the observed statistical regularities obtain. 

Coin tossing is not a process which happens only here and now but which happens at arbitrary locations and times. To explain the statistical regularities in such generality, we still need to lift the whole discussion to a universal level. The universally relevant  LLN would then have to say (very) roughly something like: 

 \medskip
\sl
\noindent For most universes in which coin tossing experiments are done, i.e., for  Liouville measure-typical such universes, it is the case that the empirical distribution of heads in long enough sequences in coin tossing experiments is approximately $1/2$.

 \medskip
\rm
The typicality assertion concerning Born's law is very analogous to this and has been proven in \cite{DGZ}. Before we turn to that we shortly look at another rather simple classical system. Everything that will be said for this example can be carried over to the case of Bohmian mechanics.

Consider an ideal gas of point particles in a rectangular box, lets say with elastic collisions of the particles at the walls. The gas is in equilibrium when the gas molecules fill the box approximately homogeneously. Most configurations (with respect to the Liouville-typicality measure) are like that, like most $0,1$-sequences have about equal numbers of heads and tails. In the course of time, there will be fluctuations of the number of molecules in a given region in the box, but those will escape our gross senses. Most configurations will stay in equilibrium over time. Now suppose we start with a gas that is occupying only one half of the box, the other half being empty. This would count as a non-equilibrium configuration. What will happen in the course of time? Well, eventually the gas molecules will fill the box approximately homogeneously. Will this relaxation happen for any possible configuration of gas that starts in one half of the box? The answer is no. For our simple example one can easily construct configurations which will never look like the equilibrium ones. There will be configurations for which it takes an enormous amount of time to evolve into ones which look like equilibrium. And some never will. Why is that important to observe? Because if one wants to make predictions about the possible behavior non-equilibrium configurations one needs to invoke typically. Namely, the idea (which is Boltzmann's insight) is that typical, i.e., \emph{most}, non-equilibrium configurations will evolve to configurations which macroscopically look like equilibrium ones. (``Most'' is again with respect to the Liouville-typicality measure, concentrated initially on the very small subset of configurations which are such that the box is only half filled.) Why? Because the equilibrium set in phase space, is so overwhelmingly larger then the tiny non-equilibrium set, so that typically trajectories will wander around and will end up in the overwhelmingly large set and stay there for  a very large time. And, as we said, there exist also atypical configurations which will not at all behave like that. That is, without typicality, we have no explanation why to expect equilibration. Having said this, we should warn the reader that this just is the physical idea behind the equilibration. To turn this into a rigorous argument is famously hard, as hard as to justify the Boltzmann equation from first principles. 

The warning in mind, we can think of  describing the transition from non-equilibrium to equilibrium also in terms of coarse-graining densities $\rho({\bf x},t)$, which are more or less smooth functions (macro-variables) on the three-dimensional physical space with variables ${\bf x}$ and which should be pictured as approximations of empirical densities.{\footnote{The empirical density is in this case given by $\rho^N_{\rm emp} (q,{\bf x}) = \frac{1}{N} \sum^N_{k = 1} \delta({\bf x} - {\bf x}_i)$, where ${\bf x}_i$ is the position of the $i$th particle. Note the analogy with the definition in the case of the coin tossing.\label{empirical}}} The uniform density i.e., $ \rho_{eq}({\bf x}) = {\textrm{const.}}$ would then be the equilibrium density. Hence, starting with a non-equilibrium density $\rho_{neq}$, it is perhaps reasonable to assume, that  $\rho_{neq}(t)\to  \rho_{eq}$ as $t$ gets large. This convergence of densities is sometimes referred to as ``mixing property" and we shall refer to this notion to mean just that: convergence of densities without reference to typicality. There have been attempts to show this. The mixing idea is presumably due to Willard Gibbs who had introduced the so-called ensemble view into statistical physics. An idea for a strategy for a ``convergence to equilibrium  proof"  was suggested by Paul Ehrenfest as is recalled on page 85 by Marc Kac in \cite{Kac} and where he refers to Ehrenfest's  attempt as an ``amusing" theorem, since convergence to equilibrium in time does not follow at all from what Ehrenfest had shown.

But even when the mixing property, i.e., convergence of densities,  were shown to be of physical relevance, the connection  with the actual configuration (i.e., the empirical distribution) would still have to be established. After all, Newtonian physics is about configurations and not densities.  
%Note also that 
In addition, by our arguments above, some non-equilibrium densities will never show the mixing property, for example densities which are concentrated on ``bad" configurations, i.e., atypical ones.

 \section{Born's Rule}
What we have said about the statistical analysis in classical physics carries over to the statistical analysis of Bohmian mechanics, where the phase space is now replaced by configuration space. Born's rule  $\rho=|\psi|^2$ is a short hand for the universal LLN for the empirical distribution $\rho^N_{\rm emp}$ of the coordinates of the particles of subsystems in an ensemble (defined similarly as in footnote \ref{empirical}). Roughly speaking, the universal LLN in the context of Born's rule says the following (for the precise formulation,  see e.g. \cite{DGZ}):

\medskip
\sl
\noindent For typical Bohmian universes hold: In an ensemble of (identical) subsystems of a universe, where each subsystem has effective wave function\footnote{Think of this wave function as the usual wave function of a system as it is used in physics courses.} $\psi$, the empirical distribution $\rho^N_{\rm emp}$ of the particles coordinates of the subsystems are  $|\psi|^2$ distributed.

\medskip
\rm
For this to hold sufficiently well, the number  $N$ of subsystems in the ensemble should be large. Note that in analogy with the coin tossing, the number $1/2$ is here replaced by $|\psi|^2$ and the sequence of length $1000$ is here the number of subsystems in the ensemble. But instead of the Liouville-measure, the typicality measure used in \cite{DGZ} is the measure $\mathbb{P}^{\Psi}(A)=\int_A |\Psi|^2(q) \rm{d} q$ ($q$ is a generic configuration space variable), where $A$ is a subset of the configuration space of the Bohmian universe and $\Psi$ is the universal wave function on that space.\footnote{As an aside, note that the typicality measure which is used in the LLN is really a member of an equivalence set of measures. That is, all measures which are absolutely continuous with respect to $\mathbb{P}^{\Psi}$ yield a LLN for Born's law.} What is special about the typicality measure $\mathbb{P}^{\Psi}$?  It is a  measure which is transported  equivariantly by the Bohmian flow. This means that it is a  typicality measure which like the stationary Liouville measure is independent of time.\footnote{It has been proven under very reasonable conditions in \cite{Ward} that this measure is unique.} 

The very nice property of the universal quantum equilibrium LLN is that it is empirically adequate. Up to date all tests affirm the empirical  validity of Born's law.

\section{Dynamical Relaxation?}
Valentini dislikes the use of typicality. Instead, he proposes  ``dynamical relaxation"  to equilibrium to explain Born's rule in the realm of Bohmian mechanics. It is however not  at all clear what is meant by ``dynamical  relaxation" and in which way  reference to typicality can be overcome. On the configurational level, i.e., on the level of empirical densities, starting in non-equilibrium 
 our  discussion of the gas in the box applies verbatim. There will always exist initial configurations of particles for which the empirical distribution will never become close to $|\psi|^2$ -- \emph{the equilibrium distribution.}  So why should we expect equilibrium then? Appealing to Boltzmann's idea, one could invoke typicality as in the case of the gas example. But as soon as one invokes typicality, there is no longer any need to invoke relaxation to begin with to explain equilibrium! Namely, most configurations will be in equilibrium most of the time and hence non-equilibrium just doesn't occur--for all practical purposes -- as established in \cite{DGZ}.
 
 Valentini also follows the Gibbs-Ehrenfest idea of mixing  and  provides an analytic argument for the convergence of densities.
 But the argument  is the direct analogue of the ``amusing" theorem proven by Ehrenfest, which ``tells us \emph{nothing} about the behavior of the non-equilibrium density $\rho$ in time" \cite{Kac}. Not to say that the connection to empirical densities needs to be established on top of that.

Hence the ``dynamical relaxation" approach turns out be neither necessary nor sufficient.

\section{Physically misguided?}
All of the quantum formalism follows from Born's rule \cite{DGZ-Op}.\footnote{This involves in some way or another an analysis of measurement situations, which in Bohmian mechanics is straightforward, and which in let's say orthodox quantum theory needs the problematical collapse of the wave function.} There is no dispute about that. Heisenberg's uncertainty follows from Born's rule. No dispute about that either. There is actually no dispute about any of the consequences which arise from or in quantum equilibrium. So what is the dispute about then? If it is about the needed reference to typicality, then that can't be because both ``approaches" need reference to typicality  anyhow for physically meaningful assertions.

What then makes the use of typicality physically misguided? Because the physical law allows for atypical universes? Because a coin tossing sequence of only heads is possible by the physical law? No argument, other than denying the physical law, can make those possibilities  impossible. Why then, don't we deny the law to make them go away?  Because by humbly looking at the facts in our world  we understand that the law-like regularities in apparently random events are in  surprising harmony with the physical law: They are typical.

\section{Acknowledgments}
W.S.\ is supported by the Research Foundation Flanders (Fonds Wetenschappelijk Onderzoek, FWO), Grant No. G066918N.

\end{document}